# On Measurement of the Spatio-Frequency Property of OFDM Backscattering

Xiaoxue Zhang*, Nanhuan Mi*, Xin He*†, Panlong Yang*, Haohua Du‡, Jiahui Hou‡ and Pengjun Wan‡
*School of Computer Science and Technology,
University of Science and Technology of China, Hefei, 230026 China
†School of Computer and Information,
Anhui Normal University, Wuhu, 241002 China
‡Department of Computer Science,
Illinois Institute of Technology, Chicago, IL 60616 USA
Email: {zxx97, nhmi95}@mail.ustc.edu.cn, {aaronhexin, panlongyang}@gmail.com,
{hdu4, jhou11}@hawk.iit.edu, wan@cs.iit.edu

*Abstract*—Orthogonal frequency-division multiplexing (OFDM) backscatter system, such as Wi-Fi backscatter, has recently been recognized as a promising technique for the IoT connectivity, due to its ubiquitous and low-cost property. This paper investigates the spatial-frequency property of the OFDM backscatter which takes the distance and the angle into account in different frequency bands. We deploy three typical scenarios for performing measurements to evaluate the received signal strength from the backscatter link. The impact of the distances among the transmitter, the tag and the receiver, as well as the angle between the transmitter and the tag is observed through the obtained measurement data. From the evaluation results, it is found that the best location of tag is either close to the receiver or the transmitter which depends on the frequency band, and the best angle is 90 degrees between the transmitter and the receiver. This work opens the shed light on the spatial deployment of the backscatter tag in different frequency band with the aim of improving the performance and reducing the interference.

## I. Introduction

The internet of things (IoT) is recognized as a new paradigm in information technology, where a large amount of devices are connected to the Internet to send the sensing data. It is expected to gain significant benefits from the IoT technology to our daily life. The key factor of pervasively deploying IoT is energy efficiency. However, reducing the energy consumption is still a critical problem and hence it strongly affect the promotion of the IoT applications. To this end, backscattering communication has been attracted attention as a promising technique to connect the IoT devices, which is energy efficient and ubiquitous.

Instead of generating their own radio frequency (RF) signals to transmit information, the nodes in the backscattering communication reflect the ambient (e.g., TV signal [1], Wi-Fi signal [2]) or dedicated signal [3] in the air[1] and modulate the signal to embed their own information bits. RF identification (RFID) is a widely used system based on backscattering, where a RFID reader generates continuous wave and the tag can reflect the signal for data transmission. Since the RF module which consumes the most portions of energy is not needed to be equipped in the backscatter node, the energy consumption is significantly reduced. On the other hand, the ambient backscatter enables ubiquitous deployment of the IoT due to the fact that no special infrastructure is required. Therefore, the IoT applications with low power constraints, such as implantable sensors, home appliances, could benefit from the backscatter communication.

Similar to RFID system, a specialized hardware is required in dedicated backscatter systems to generate excitation signal and decode the reflected signal, which brings additional cost and is not convenient to deploy anywhere. Recent work on backscatter the existing Wi-Fi and Bluetooth radios from commodity devices, such as Wi-Fi backscatter [4], BackFi [2], FreeRider [5], HitchHike [6] and implicit Wi-Fi backscatter [7], has reduced the requirement for specialized hardware. The passive tag in BackFi system is able to communicate with Wi-Fi access points (AP) using standard Wi-Fi packets as excitation signals. FreeRider system can simultaneously use 802.11g/n Wi-Fi, ZigBee and Bluetooth to enable data communication. HitchHike system uses 802.11b Wi-Fi as the excitation signal and performs codeword translation to achieve backscatter communication using off-the-shelf Wi-Fi routers. The implicit Wi-Fi backscatter system adopts a flicker detector to enable per-symbol and in-band backscatter by utilizing residual channel of Wi-Fi packets. Besides these prototype systems, several theoretical analyses, such as [8], [9], and link layer design [10] are also presented.

---

[1]Hereafter, these signals are referred to as excitation signals in backscatter system.

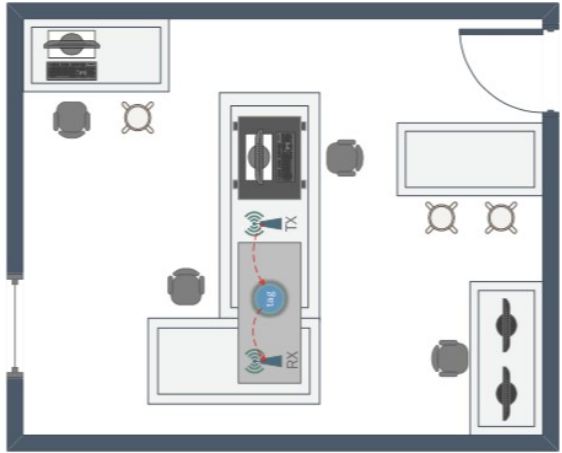

Fig. 1. The working environment of conducting measurements.

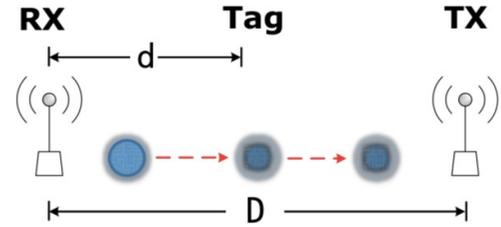

Fig. 2. Experiment scenarios of evaluating backscattering OFDM signals performance change with distance.

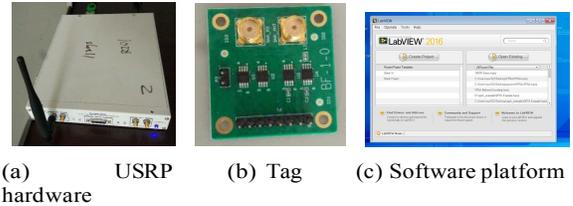

(a) USRP hardware    (b) Tag    (c) Software platform

Fig. 3. The software and hardware platforms of performing measurement.

However, these state-of-the-art work only focus on the design of the transmission techniques. It is notably important to perform a series of measurements on backscatter orthogonal frequency-division multiplexing (OFDM) signal in terms of received signal strength for different scenarios. Based on the measurement data, it could find the opportunities to better deploy the passive tags for improving the quality of the reflected signal and reducing the interference between tags. Therefore, we present this work, the first measurement campaign on backscatter OFDM signals. We evaluate the impact of the distances and the angles among the transmitter, tags and the receiver, on the received signal strength to find the spatial-frequency property of backscatter communications. It is observed from the measurements that the best location of tag is close to the receiver or the transmitter, and the best angle is 90 degrees.

The rest of this paper is organized as follows. Section II introduces the background and the motivation of backscatter communication. The experiment setup of performing measurement is detailed in Section III. The evaluation results and their discussions are presented in Section IV. We give some concluding statements in Section V. [t]

## II. Background and Motivation

In this section, we first give an overview of the backscatter systems. A backscatter system generally consists of three components: a transmitter, a tag and a receiver. The transmitter generates an excitation signal as the plug-in device in passive Wi-Fi design, as well as serves as a charging infrastructure to offer the required power of activating the circuit of the tag. For simplicity, the signal can be represented as a sinusoidal signal, denoted as $\sin(2\pi f_c t)$. The backscatter tag is composed of an antenna and a micro-controller, which controls the single pole double throw (SPDT) RF switch group to reflect the incoming signals or absorb the energy of the signals. By changing the impedance of the antenna, the tag can switch its states between reflecting and non-reflecting. In order to control the impedance of the antenna, the micro-controller generates a square wave at a frequency of $\Delta_f$. Since the first harmonic of a square wave is also a sinusoid signal based on the well-known Fourier analysis, when applying to incoming sinusoidal signal, a frequency shift $\Delta_f$ is achieved in the backscatter link. Thus, we can simplify the process of square wave to a sinusoid signal using the first harmonic, as $\sin(2\pi \Delta_f t)$. Consequently, the process of backscattering can be represented by the product of the aforementioned two sinusoidal signals, which is given by $2\sin(2\pi f_c t)\sin(2\pi \Delta_f t) = \cos(2\pi(f_c - \Delta_f)t) - \cos(2\pi(f_c + \Delta_f)t)$. As a process for the received backscattered signals, a receiver node turns its center frequency to one of the shifted signal, which is $f_c - \Delta_f$ and $f_c + \Delta_f$ in our configuration. It should be emphasized here that the purpose of the frequency shift is to eliminate the interference from the main link to the backscatter link, as the signal strength of the main link is significantly stronger than that of the backscatter link. However, the frequency shift brings additional power consumption to generate the square wave at the tag, and results in interference to the shifted frequency band.

## III. Experiment Setup

OFDM. Widely applied commodity Wi-Fi 802.11g/n adopts the modulation scheme called orthogonal frequency-division multiplexing. Its main idea is to divide the channel into many orthogonal sub-channels before performing narrow-band modulation and transmission on each sub-channel. OFDM has been developed into a popular scheme for wideband digital communication, adopted in many communication standards such as digital television and audio broadcasting, digital subscriber line internet access, wireless networks, power line networks, and 4G mobile communications. The main advantages of OFDM are high bit rate, robust against narrow-band co-channel interference and high spectral efficiency. Considering the extensive use of the OFDM signal and its merits, we apply OFDM signal as original signal to explore the factors that influence the power of backscattering signal.

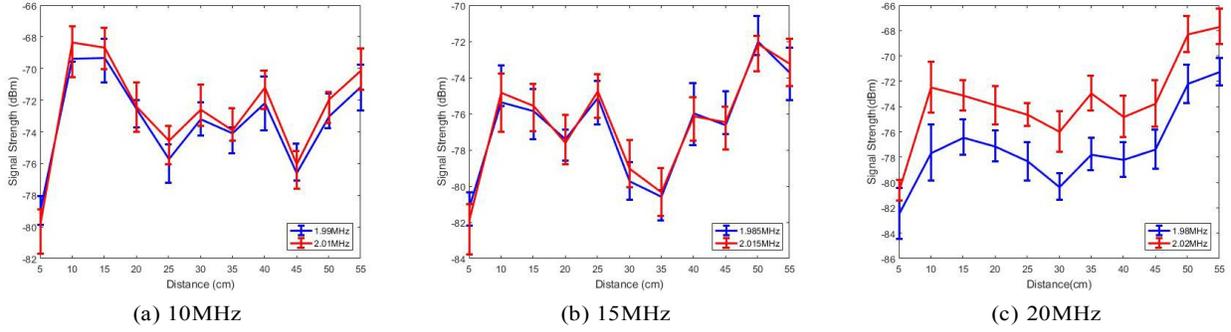

Fig. 4. Received backscatter signal strength when tag moves along the line connecting the transmitter and receiver.

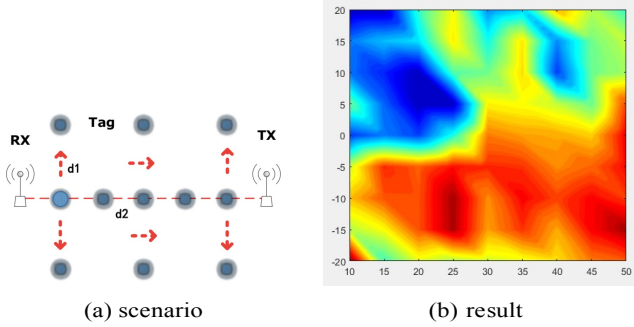

(a) scenario  (b) result

Fig. 5. Experiment scenarios of evaluating backscattering OFDM signals performance change with distance.

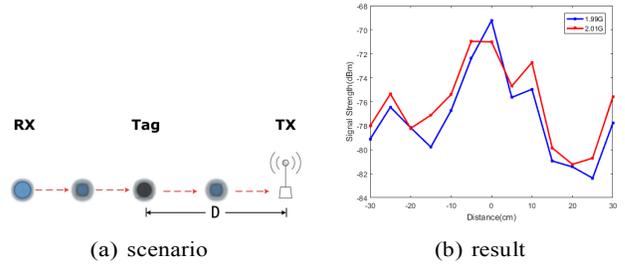

(a) scenario  (b) result

Fig. 6. Received backscatter signal strength when tag moves along the line connectiong the transmitter and receiver.

Figure 1 shows the experiment scenario of communication ranges in a 8 ×9 m indoor area. It adopts passive wireless backscattering devices and leverages wireless AP as their powering infrastructure and excitation signal source. The system is implement in the Labview and FPGA platform using the USRP RIO devices, as shown in Fig. 3.

It should be mentioned here that the aim of our experiments is to find the spatial-frequency property from the measurement data, i.e., how the spatial deployment using different frequency shift affect the backscatter signal?

A. Scenario 1-Straight Line

In many scenarios of backscatter system in practice, the size of the receivers are much larger than that of the tags, and those tags without RF module are easier to move. To this end, we explore the effect of the position of the tag on the power of the received reflected signal while the positions of the transmitter (TX) and the receiver (RX) are fixed.

In our experiment, the OFDM transmitter and the receiver have fixed positions, and the tag moves along a straight line to increase the distance away from the receiver.

B. Scenario 2-Two-Dimensional Plane

Consider the situation in a practical backscatter communication system, tags can move beyond a straight line between TX and RX. For a better intuition, we extend the position in the Scenario 1, to explore the signal strength when the tag moves in a two-dimensional plane constructed by the RX and the TX.

In this setup, the OFDM transmitter and the receiver have fixed positions, and the tag moves in TX-RX platform.

C. Scenario 3-Impact of the Angle

A more practical scenario is that, the TX and tags are deployed with fixed positions respectively, while the RX moves around with users. For instance, a RX can be a mobile phone, a wearable device etc. Therefore, in addition to the TX and RX fixed scenario, we further evaluate our system in the TX and tag fixed scenario.

In this scenario, we conduct two experiments to evaluate the impact of the angle. In the first one, the OFDM transmitter and the tag have fixed positions, and the receiver moves along a straight line to increase the distance away from the tag, which illustrates the impact of the distance between tag and receiver. In the second experiment, the distance between the receiver and tag is fixed, while the receiver is placed at different directions.

## IV. Evaluation

In this section, we evaluate the performance of backscattered OFDM signals in an indoor environment in our lab with the experimental scenarios described above. We examine how the backscatter signal strength is affected by the shifted frequency $\Delta_f$, the distances (both TX- to-tag distances and tag-to-RX distances) and the angle between the RX and the TX, when the carrier frequency of original OFDM is 2 GHz. We also study the interference between two tags. The labels of scenarios are the same as those in section I.

### A. Scenario 1

We place the OFDM transmitter and the receiver separated by 60 cm as shown in Fig. 2. We move the tag from the receiver to the transmitter at a 5 cm interval. Since the shifted frequency $\Delta_f$ of the tag may also affect the strength of backscattered signals, three types of shifted frequency are used, including 10 MHz, 15 MHz and 20 MHZ.

Figure 4 shows how the received backscatter signal strength varies with respect to the distance between RX and tag, when tag moves along the line connecting RX and TX. As shown in Fig. 4, for all types of shifted frequencies of the tag, backscatter signal strength shows a general trend of decreasing from 10 cm and then increasing according to the Friis path loss model. The mid-point between the transmitter and the receiver has the lowest strength, which can be straightforwardly obtained using the path loss model and letting the first-order derivative equal to 0.

However, the point at 5 cm reverses the general trend. Based on the relationship between the frequency $f$ and the wavelength $\lambda$, as $\lambda = v/f$ with $c$ being the speed of the light, we have $\lambda = \frac{3 \times 10^8}{2 \times 10^9} = 0.15$. As the half wavelength of carrier signals at $f = 2$ GHZ is 7.5 cm, the possible reason of the reversing point is that the distance is less than the half of the wavelength. Comparing the results of different types of tags with different shifted frequencies $\Delta_f$, it is found that when the frequency of tag increases, the difference of signal strength between the two shifted backscattering signals increased.

### B. Scenario 2

In this experiment, we still place the OFDM transmitter and receiver separated by 60cm, and move the tag in TX-RX platform as illustrated in Fig. 5(a) shows. In this case, we only consider the situation that the shifted frequency of the tag is set at 10 MHz. Fig. 5(b) demonstrates the signal strength in this two-dimensional plane, through this, a best position to receive the reflected signal can be found, which is $(25, -10)$.

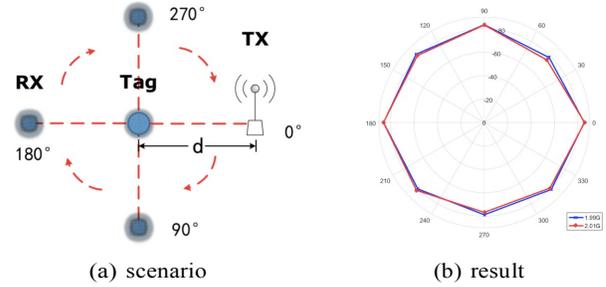

(a) scenario  (b) result

Fig. 7. Received backscatter signal strength when tag moves along the line connectiong the transmitter and receiver.

### C. Scenario 3

We now evaluate the impact of different positions of the receiver. In this scenario, we fix the transmitter and the tag with the distance of 30cm. As shown in Figure 6(a), the RX is moved from the position A to the position B. Figure 6(b) shows the trend of increased signal strength with receiver moving close to the tag, and decreased signal strength when moving away. In brief, we can find that the received signal increases as the receiver gets closer to the tag. In addition, Fig. 7 illustrates the received signal strength with respect to the direction of RX. As shown in Fig. 7(a), the distance between the receiver and the tag is 30 cm, and we change the direction of RX in circle centered on tag. The result of Fig. 7(b) shows that the signal strength is the highest at the angel of 90 degree between the RX and the TX.

### D. Discussion

From the above measurements, it is interesting that the shifted frequency $\Delta_f$ of the tag also affected the received signal strength from the backscatter link. Furthermore, the impact of the distances among RX-tag-TX is not well consistent with the theoretical results which is shown in Fig. 8. The reason is that the theoretical analysis does not fully take the impendence of the hardware design and the shifted frequency into account. However, the measurement data offer the opportunities to design the spatial deployment scenario in backscatter communications for different IoT applications.

## V. Conclusion

This paper evaluated the spatial-frequency property of the OFDM backscatter which consider the distance and the angle in different frequency bands. We took three scenarios for performing measurements to measure the received signal strength of the backscatter link. The impact of the distances among the transmitter, the tag and the receiver, as well as the angle between the transmitter and the tag is observed through the obtained measurement data. From the evaluation results, it is found that the best location of tag is either close to the receiver (shifted 10 MHz) or the transmitter (shifted 15 and 20 MHz) which depends on the frequency band, and the best angle is 90 degrees. This work opens the shed light on the spatial deployment of the backscatter tag in different frequency band with the aim of improving the performance and reducing the interference. As a future study, the measurements of concurrent multi-tag backscatter communication is ongoing.

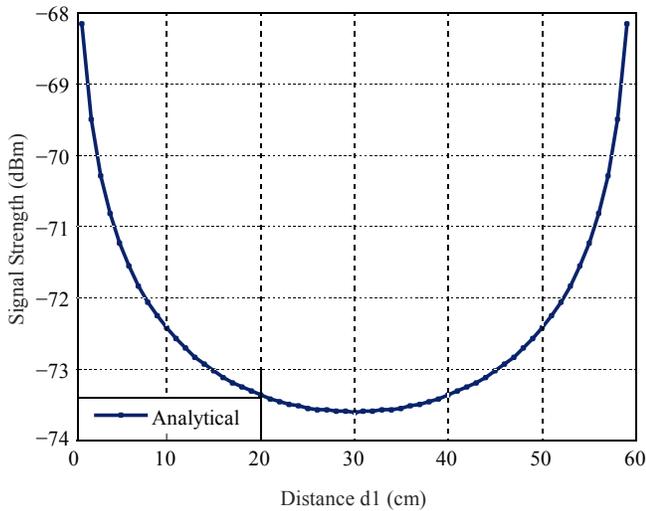

Fig. 8. The analytical signal strength of backscatter link using the path-loss model.

## Acknowledgement

This work was supported in part by the National Natural Science Foundation of China under Grants No. 61702011, and in part by the National Science Foundation of USA under grants CNS-1526638.